%% file: main.tex
\documentclass[conference,a4paper]{IEEEtran} 

\IEEEoverridecommandlockouts

 \IEEEpubid{\makebox[\columnwidth]{978-1-7281-3549-6/19/\$31.00~\copyright{}2019
 IEEE \hfill} \hspace{\columnsep}\makebox[\columnwidth]{ }}

\input{share/tex/setupIEEE.tex}

\begin{document}

\title{Visualizations for Understanding SoC Behaviour \\
\thanks{This project is supported by the Engineering and Physical Sciences
Research Council (EP/I028153/ and EP/L016656/1);
the University of Bristol and UltraSoC Technologies Ltd.
Supervised by Dr Jose Nunez-Yanez and Professor Kerstin Eder.}
}

\author{
    \IEEEauthorblockN{Dave M\textsuperscript{c}Ewan}
    \IEEEauthorblockA{\textit{Centre for Doctoral Training in Communications} \\
    \textit{University of Bristol}\\
    Bristol, UK \\
    dave.mcewan@bristol.ac.uk}
\and
    \IEEEauthorblockN{Marcin Hlond}
    \IEEEauthorblockA{\textit{Director of System Engineering} \\
    \textit{UltraSoC Technologies Ltd}\\
    Bristol, UK \\
    marcin.hlond@ultrasoc.com}
\and
    \IEEEauthorblockN{Dr Jose Nunez-Yanez}
    \IEEEauthorblockA{\textit{Dept. of Microelectronics} \\
    \textit{University of Bristol}\\
    Bristol, UK \\
    j.l.nunez-yanez@bristol.ac.uk}
}

\maketitle

\begin{abstract}
This paper introduces a novel method of analysis for \gls{soc} development
building upon commonly used tools and techniques to approximate and automate
the human process of investigation.
Knowledge of the interactions between components within a \gls{soc} is
essential for understanding how a system works so the presented method provides
a way of visualizing these interactions.
The mathematical basis for the method is explained and justified, then the
method is demonstrated using two representative case studies.
Visualizations from the case studies are used to exhibit the usefulness of the
method for system optimization, monitoring, and validation.
\end{abstract}


\section{Introduction} 
\label{sec:introduction}
The \gls{soc}s comprising modern silicon products are often built with the
result of lifetime's of work by hundreds of engineers which makes it all but
impossible for a single systems architect to have a complete understanding of
every part in a design.
This means that while at first glance a system may appear to be functioning,
unforeseen behaviours may appear within the interactions between system
components, possibly leading to undesirable behaviour such as reduced
performance, increased energy usage, information leakage, or unexpected
susceptiblity to faults.
\gls{ust} is a silicon IP supplier specializing in embedded analytics
which uses highly configurable monitoring components to address these issues.
Analysing so much data using traditional methods such as assertions is difficult
and time consuming due to systemic complexity and dynamic behaviour.
In this research we propose a visual tool and mathematical framework that can
help to understand these behaviours and build upon the instrumentation
technology developed by project partner \gls{ust}.
Two cases studies are used to illustrate the methodology:
A synthesizable model of a simple \gls{soc}, and a complex \gls{soc} with
lightweight software instrumentation enabled by the use of \gls{ust} tools.

The main contributions of this paper are:
1) A mathematical framework for approximating the human process of
investigation for binarized time series \gls{soc} data.
2) A novel visualization technique for behavioural relationships.


\section{Previous Work} 

An examination of currently available hardware and low-level software
profiling methods is given by Lagraa\cite{LagraaThesis} which covers well known
techniques such as using counters to generate statistics about both
hardware and software events -- effectively a low cost data compression.
Lagraa's thesis is based on profiling \gls{soc}s created specifically on
Xilinx MP-SoC devices, which although powerful, ensures it may not be applied to
data from other sources, such as in post-silicon.
Lo et al\cite{MiningPastTemporalRules} described a system for describing
behaviour with a series of statements using a search space exploration process
based on boolean set theory.
While this work has a similar goal of finding temporal dependencies it is
acknowledged that the mining method does not perform adequately for the very
long traces often found in real-world \gls{soc} data.
Another limitaton here is our receptiveness to information in the form of long
lists of statements versus a visual representation.
Ivanovic et al\cite{TSAnalysisPossApp} review time series
analysis models and methods where characteristic features of economic time
series are described such as high auto-dependence and inter-dependence, high
correlation, non-stationarity, and drawn from noisy sources.
\gls{soc} data is expected to have these same features, together with
full binarization and much greater length.
Explainability is a key requirement to understanding so related approaches such
as the use of \gls{nn}s has been avoided at this stage although these may be
useful for higher level analysis.


\section{Methodology} 
\label{sec:method}
This methodology has been designed to approximate and mimic the process of an
experienced \gls{soc} engineer trying to understand how waveforms are related
to each other.
It is assumed that the measurements are taken at discrete times $t$, often
referred to as a number of clock cycles, and that all values are binary,
$f_i(t) \in \{0, 1\}$.
Additionally it is assumed that data for every time is able to be recorded, or
accurately inferred, which depends on the changes in measurement values to be
sparse for storing data at a physically feasible rate.

The first approximation utilized is the application of a bell-shaped windowing
function $w$ which is similar to how we focus
attention on the centre of a time period $t \in [u,v)$.
A power-of-sine window windowing function $w$ is used to create a weighted
average of each measurement giving the expected value.
\begin{align}
\label{eq:win_powsin}
w(t) &=
    \begin{cases}
    {\sin^\alpha}{\left( \frac{t\pi}{v-u-1} \right)} &: t \in [u,v)
    \\
    0 &: \text{otherwise}
    \end{cases}
\\
\label{eq:def_Ex}
\sEx[f_i] &= \frac{1}{\sum w}
    \mathlarger{\sum}_{t \in [u,v)} w(t) * f_i(t)
\quad \in [0,1]
\end{align}

Bayes theorem in \Fref{eq:bayes} and the definition of independence in
\Fref{eq:independence} to allow the conditional expectation and a
measure of dependency, $\sDep$, to be calculated.
In order to reduce the amount of information stored, a threshold
is introduced which approximates the process of putting relationships in
natural language form.
\begin{align}
\label{eq:bayes} 
\Pr(X|Y) &= \frac{\Pr(Y|X) \Pr(X)}{\Pr(Y)};
    \quad \Pr(Y) \neq 0
\\
\label{eq:def_cEx}
\sEx[f_x|f_y] &= \frac{\sEx[f_x * f_y]}{\sEx[f_y]};
    \quad \sEx[f_y] \neq 0
\\
\label{eq:independence}
X \indep Y &\iff \Pr(X) = \Pr(X|Y)
\\
\nonumber
\text{let} \quad \varphi &= \frac{{\sEx}{\left[ f_x | f_y \right]} - {\sEx}{\left[ f_x \right]}}
                                 {{\sEx}{\left[ f_x | f_y \right]}}
                          = 1 - \frac{{\sEx}{\left[ f_x \right]} {\sEx}{\left[ f_y \right]}}
                                 {{\sEx}{\left[ f_x * f_y \right]}}
\\
\label{eq:def_Dep}
\sDep(f_x,f_y) &:=
    \begin{cases}
        \varphi &: 0 \leq \varphi
        \\
        0 &: \text{otherwise}
    \end{cases}
\end{align}

$\sCov$ is a measure of covariance as shown in \Fref{eq:def_Cov}.
Here the use of a weighted average and a threshold function approximate the
human processes of focusing attention and discarding pairwise correlations
which are insignificantly small or negative.
\begin{align}
\label{eq:covariance}
\cov(X, Y)
&= {\Ex}{\left[ XY \right]} - {\Ex}{\left[ X \right]}{\Ex}{\left[ Y \right]}
\\
\label{eq:covariance_limits}
X,Y \in [0,1] &\implies \frac{-1}{4} \leq \cov(X,Y) \leq \frac{1}{4}
\\
\nonumber
\text{let} \quad \varphi &=
    4 \Big(
           {\sEx}{\left[ f_x * f_y \right]} -
           {\sEx}{\left[ f_x \right]}{\sEx}{\left[ f_y \right]}
      \Big)
\\
\label{eq:def_Cov}
\sCov(f_x, f_y) &:=
    \begin{cases}
        \varphi &: 0 \leq \varphi
        \\
        0 &: \text{otherwise}
    \end{cases}
\end{align}
The measures $\sDep$, and $\sCov$ are symmetric, i.e. $\sDep(X,Y)=\sDep(Y,X)$,
share the same codomain $[0,1]$ and operate in the same domain $[0,1]$ which
allows their output to form new measurements for a meta-analysis.

\begin{figure}[t] 
\centering
\includegraphics[width=1.0\linewidth]{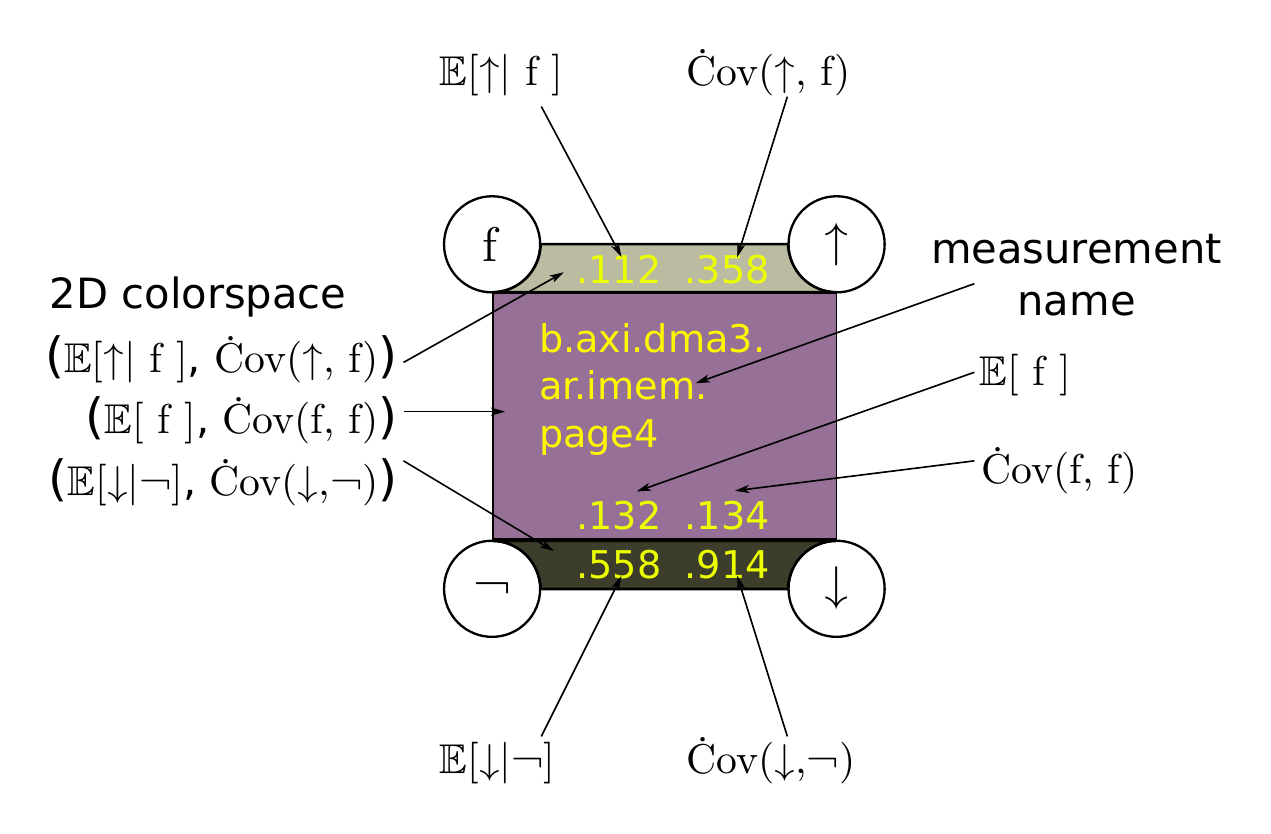}
\caption{Mockup of the representation of a single measurement.
Colours from the 2D colourspace allow the viewer to quickly estimate the values
of 6 values of interest.
\label{fig:own_net}}
\end{figure}

Each measurement lends itself to implied relationships, e.g.
``X \textit{high} leads or \ldots'' or ``X \textit{rising} leads to \ldots''.
This methodology is focused on binary measurements so four implied measurements
are considered:
\begin{enumerate}
\item Measurement $f(t) \in [0,1]$.
\item Reflection, ${\neg}(t) := (1-f)$.
\item Rising edge, ${\uparrow}(t) := {\max}\big(0, f(t)-f(t-1)\big)$.
\item Falling edge, ${\downarrow}(t) := {\max}\big(0, \neg(t)-\neg(t-1)\big)$.
\end{enumerate}
Using these four implied measurements for each real one means that the use of
thresholds to effectively discard negative dependencies and covariances does
not miss significant relationships, but puts them into the more natural form
e.g. ``X \textit{low} \ldots'' vs ``X \textit{not high} \ldots''.
The technique used to visualize some information about each measurement in a
single time window combines these into a quad, as shown in \Fref{fig:own_net}.
This allows some important information to be gleaned from just the colour of the
quad sections, allowing some usage even in a blurred, low-resolution
or faraway view where the text is unclear.

\begin{figure}[t] 
\centering
\includegraphics[width=.50\linewidth]{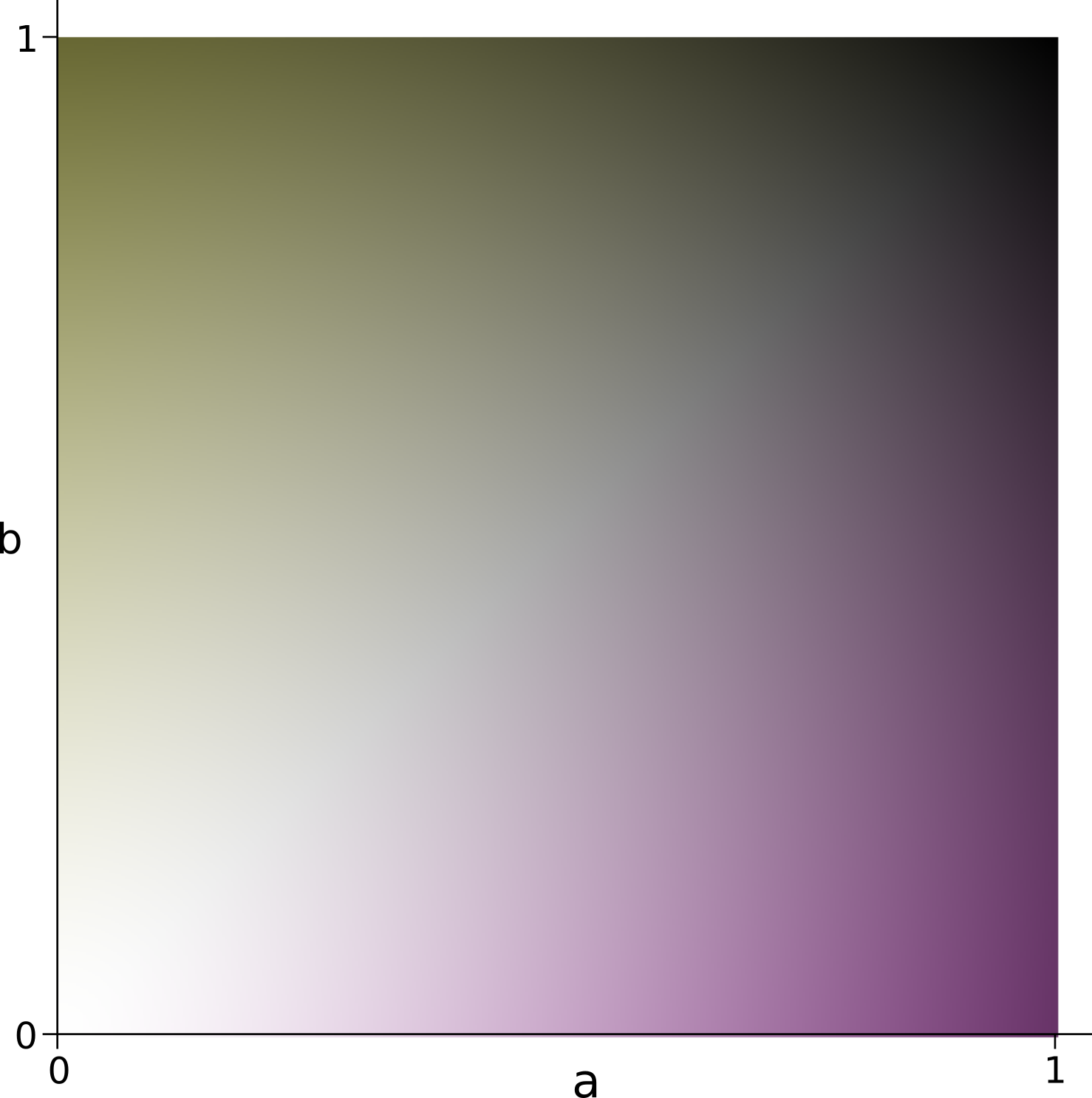}
\caption{Colourspace for visualization of 2-dimensional bounded values described
by \Fref{eq:cs0_theta} thru \Fref{eq:cs0_B}.
\label{fig:cs0}}
\end{figure}

A novel visualization has been developed in order to represent data points
in the space $[0,1]^2$ which allows the viewer to quickly determine the rough
location of a point from its colour.
\Fref{eq:cs0_theta} thru \Fref{eq:cs0_B} show the mapping from $[0, 1]^2$
to 8-bit \gls{rgb} values as depicted in \Fref{fig:cs0}.
This colourspace displays equally well on screen and printed paper and looks
similar to people with protanopia, the most common form of colourblindness.

\begin{figure}[t] 
  \centering
  \begin{subfigure}[t]{0.5\textwidth}
  \includegraphics[width=1.0\linewidth]{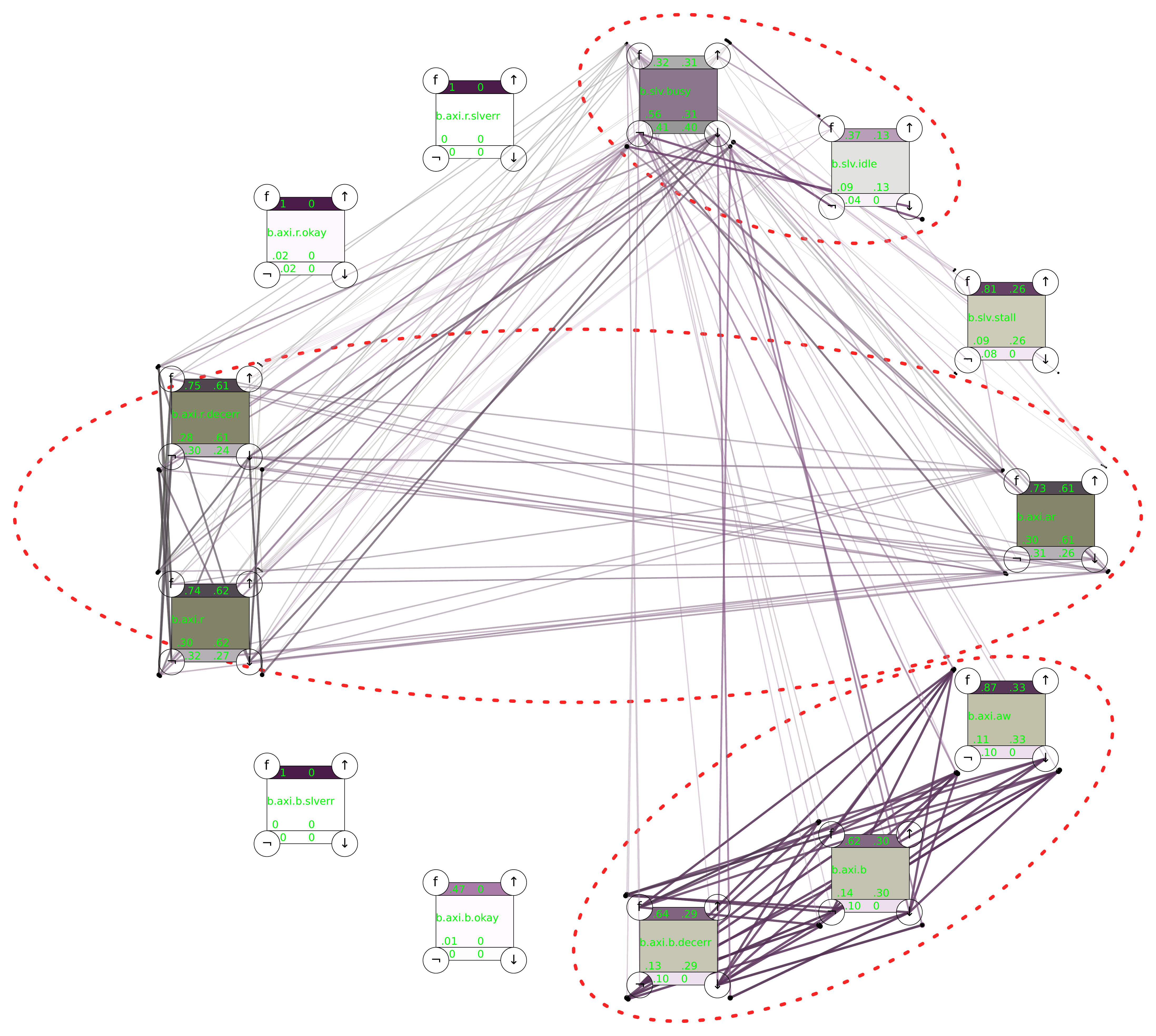}
  \caption{Plot showing measurement relationships from probsys.
  Interactions are clearly visible from the more densely connected nodes.
  Clusters of relations are outlined in red.
  Top-right \texttt{busy} and \texttt{idle} are strongly related.
  Middle \gls{axi} read request (\texttt{AR}) is strongly related to read reply
  (\texttt{R}) and read-decode error.
  Bottom-right  \gls{axi} write request (\texttt{AW}) is strongly related to
  write reply (\texttt{B}) and write-decode error.
  Read measurements are related to write measurements via \texttt{busy}.
  \label{fig:eg_probsys_net_annotated}}
  \end{subfigure}%

  \begin{subfigure}[b]{0.5\textwidth}
  \centering
  \includegraphics[width=0.5\linewidth]{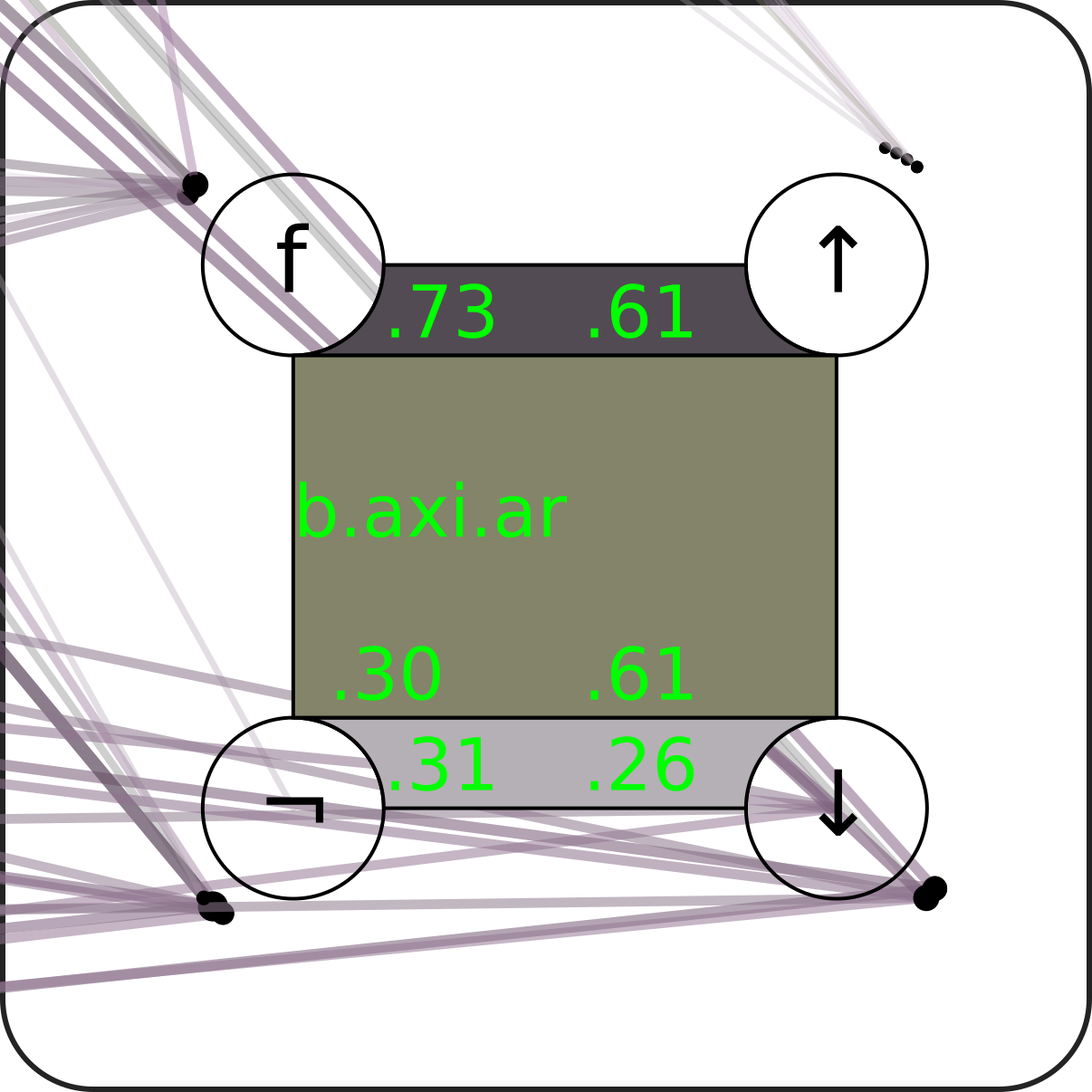}
  \caption{Closeup view of rightmost measurement node from the same plot.
  Each edge is drawn as a line from a black dot just outside one of the quad's
  corners, to the centre of another quad corner.
  Location of the dot depends on the value of $\delta$.
  The size of the dots, weight, and thickness of the drawn edges are further
  visual indicators of the $\sDep$ and $\sCov$ values.
  \label{fig:eg_probsys_net_closeup}}
  \end{subfigure}%

  \caption{
  \label{fig:eg_probsys_net}}
\end{figure}

When looking for a pairwise relationship it is necessary to look slightly
forward or backward in time to find a result such as ``X is likely 5 cycles
after Y''.
The notation $f_{i_{\langle \delta \rangle}}(t) := f_{i}(t + \delta)$ has been
used to represent the notion of measurement $i$ being shifted by $\delta$
cycles.
A link between two measurements $f_x$ and $f_{y_{\langle \delta \rangle}}$
is said to be significant when both
$\sDep(f_x,f_{y_{\langle \delta \rangle}})$ and
$\sCov(f_x,f_{y_{\langle \delta \rangle}})$ are greater than zero.
For a high level understanding, knowing that a significant link exists is
more important than knowing the exact values of dependency and covariance,
and a rough estimate may often be good enough.
By arranging all measurements in a circle and drawing edges, again using the
colourspace described above, a network of behavioural relationships is formed, as
shown in the diagrams generated from case studies \Fref{fig:eg_probsys_net}
and \Fref{fig:eg_tinn_net}.
This arrangement shows behaviour as connections of a graph with node attributes
providing a summary of the measured behaviour in an easily
digestible manner, ultimately saving engineering time by automating a large part
of the human analysis process.
\begin{align}
\label{eq:cs0_theta}
\theta &= \left(1 - \frac{\sqrt{a^2 + b^2}}{\sqrt{2}}\right)^\gamma
\\
\label{eq:cs0_phi}
\phi &= \arctan{\frac{b}{a}}
\\
\label{eq:cs0_R} 
\text{red} &=
    \lfloor 255 \times \theta \rfloor
\\
\label{eq:cs0_G} 
\text{green} &=
    \lfloor 255 \times \theta^{\max(0,\ \frac{\pi}{4} - \phi) + 1} \rfloor
\\
\label{eq:cs0_B} 
\text{blue} &=
    \lfloor 255 \times \theta^{\max(0,\ \phi - \frac{\pi}{4}) + 1} \rfloor
\end{align}


\section{Case Study 1} 
\label{sec:probsys}
The first experiment named \textit{probsys} is based on a system consisting of
a single \gls{axi}\cite{AMBA4} master communicating with a single \gls{axi}
slave.
To give some familiar context three additional binary states are measured on
the slave component (\texttt{busy}, \texttt{stall}, and \texttt{idle}).
A simulation is run to produce a \gls{vcd} file containing measurment data.
The rates at which transactions are made on the five \gls{axi} channels
(\texttt{AW}, \texttt{W}, \texttt{B}, \texttt{AR}, \texttt{R}) are controlled
with a probability distributon via fixed inputs.
On the source side stall this means dropping \texttt{*READY} and on the
destination side this means dropping \texttt{*VALID}.
Observing the measurements in a waveform viewer in order to validate system
behaviour is not trivial due to the density and format of information.
Some expected behaviours may be expressed as formal properties and proven.
E.g. Assertions may be used to check that \texttt{busy} and
\texttt{idle} are never high at the same time.
However, listing and forming all of these properties is time consuming and gives
no hints to what behaviours a user might have forgotten to specify.

This visualization can be inspected interactively and intuitively to find more
information at deeper levels such as the precise conditional probability,
dependency, and covariance of each edge.
Even at the static and low resolution of printed paper, much useful
information is immediately available with its significance denoted by the
darkness of the ink.
The edges correctly identify the measurements which are expected to be related
to each other, e.g. \gls{axi} read replies strongly related to \gls{axi}
read requests.
It can be quickly seen that the read channels are not interacting with the
write channels, validating (or invalidating) expectations.
This gives the system designer some confidence that the read and write parts of
the system have not been accidentally linked through some unexpected mechanism.
This example is analogous to the many use-cases where it is desired to visualize
the cooperation between channels, non-cooperation between channels,
or iteraction with a side-channel like \texttt{busy}.


\section{Case Study 2} 
\label{sec:tinn}
A system which is more representative of the complex \gls{soc}s used today is
required to demonstrate the method in a real-world application.
The application chosen for this experiment is a software-based \gls{nn}
performing handwritten-digit recognition, running on a standard prototyping
FPGA \gls{soc}\cite{UltraSocTaygete} with two
RISC-V processors and communicating with the outside world over USB.
The C software of 17 functions running on 2 processors, is based on
Tinn\cite{tinn}\cite{semeionHandwrittenDigits} 
and modified such that the smaller processor \texttt{ACPU} collects batches and
passes them to the processor which has the full floating point unit
\texttt{SCPU}.

\begin{figure}[t] 
  \centering
  \begin{subfigure}[t]{0.5\textwidth}
  \includegraphics[width=1.0\linewidth]{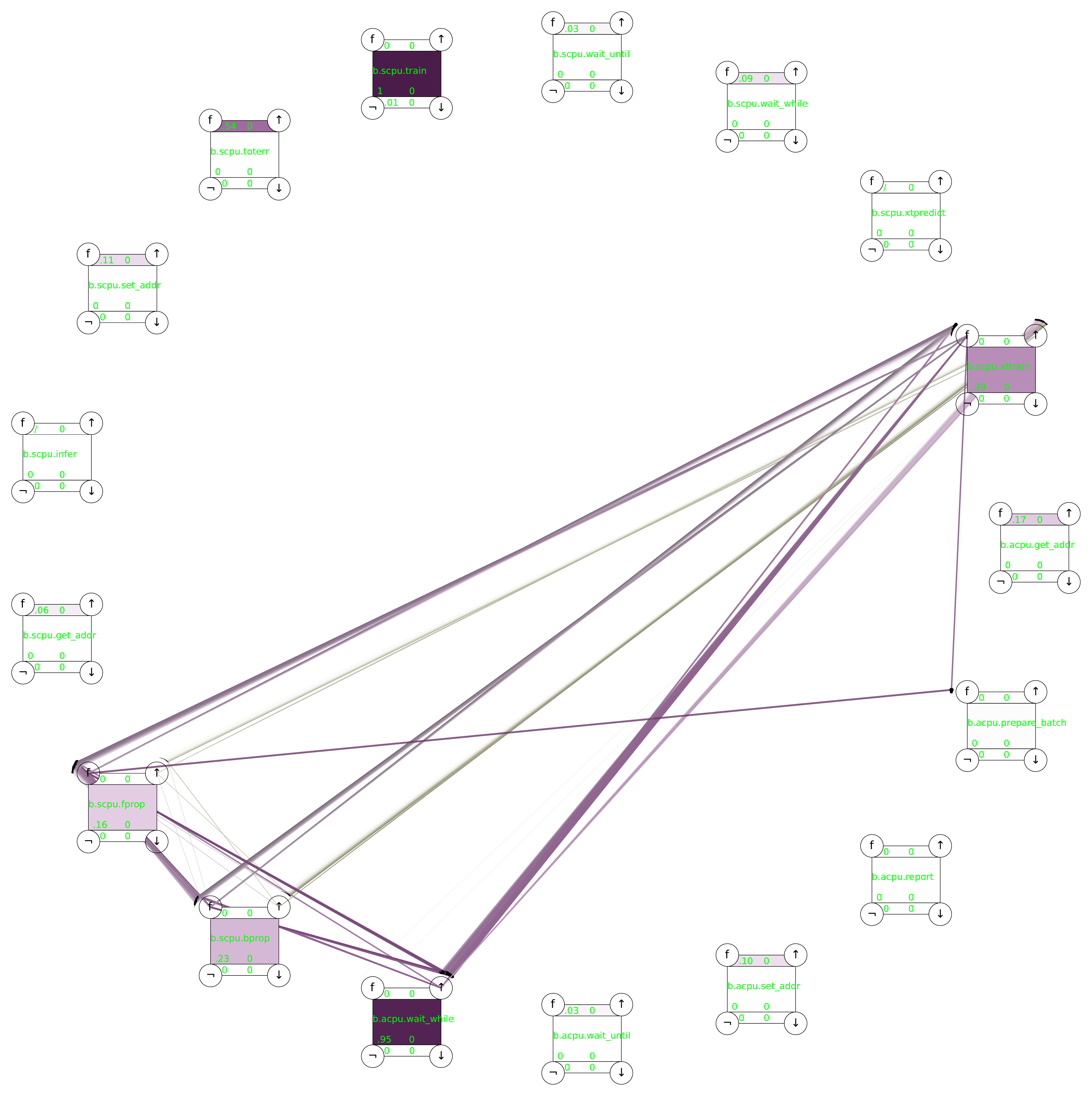}
  \caption{Behaviour graph of tinn during training phase.
  \label{fig:eg_tinn_net_train}}
  \end{subfigure}%

  \begin{subfigure}[b]{0.5\textwidth}
  \includegraphics[width=1.0\linewidth]{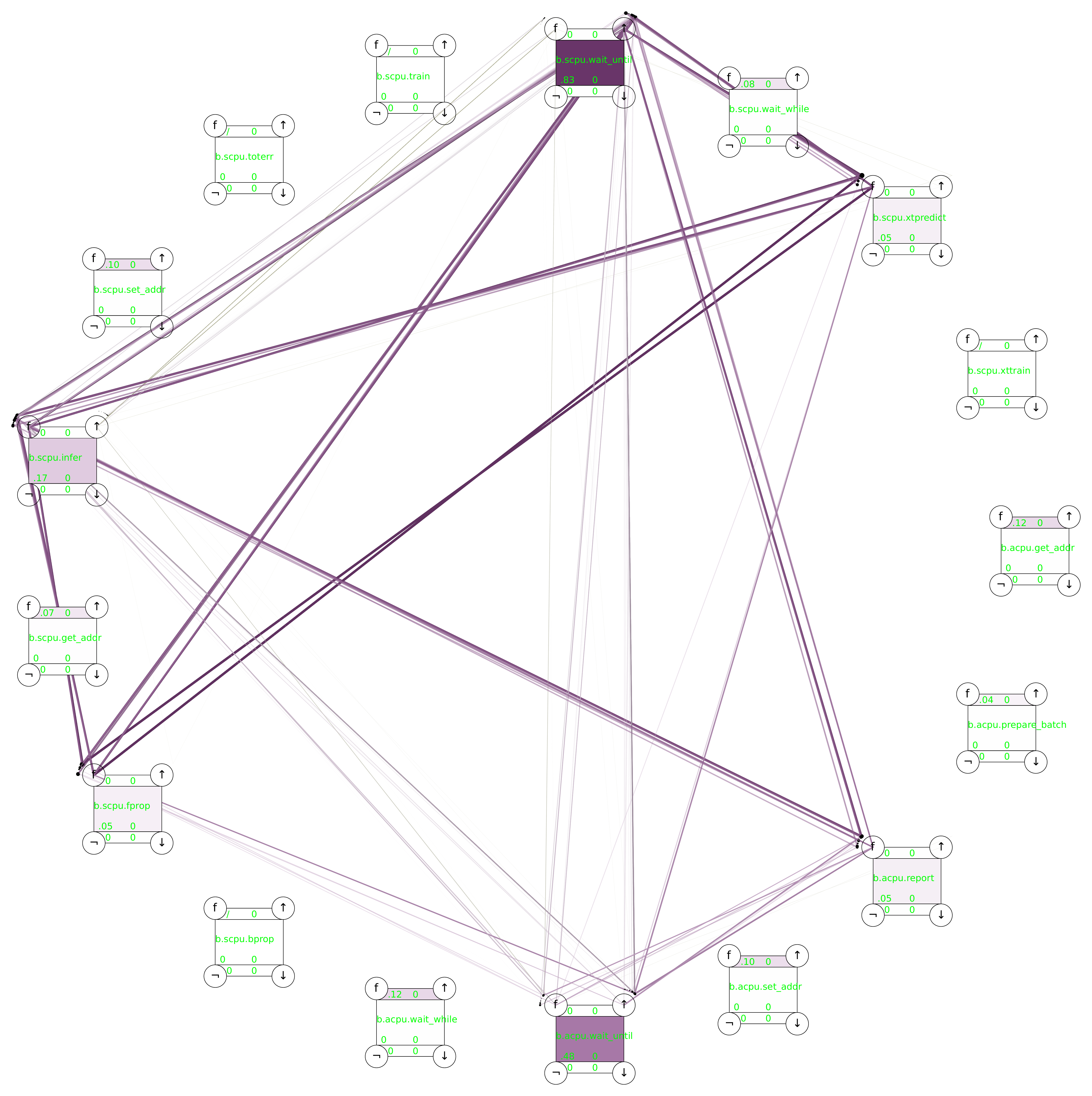}
  \caption{Behaviour graph of tinn during inference phase.
  \label{fig:eg_tinn_net_infer}}
  \end{subfigure}%

  \caption{The difference in behaviour graphs is clear, allowing a viewer to
  get a fast and useful overview of the main component interactions and how
  they change over time.
  The pattern of edges stays fairly constant in each phase, with the transition
  marked by many edges fading in and out as the `training' pattern morphs into
  the `inference' pattern.
  Immediately obvious is that most nodes are unconnected, and closer inspection
  of the connected nodes reveals that the connections are as one would expect
  from a \gls{nn} application.
  \label{fig:eg_tinn_net}}
\end{figure}

Ease of instrumentation is essential for real world systems as non-standard
modifications can alter observed behaviour is subtle but important ways.
The only modifications required are an additional GCC flag in the compile
stage (\texttt{-finstrument-functions}), and to optionally disable
instrumentation on uninteresting functions\cite{UltraSocSI}.
Network diagrams were generated using the methodology
described above, two of which are shown in \Fref{fig:eg_tinn_net}.
Since each of these diagrams corresponds to a single time window it is natural
to view them in sequence like a movie.
Using the \gls{svg} image format also allows the diagrams to be browsed and
examined interactively to extract more detail and exact values as desired.
Nodes with darker colours in the centre bands indicate more time spent
in those functions.
E.g. \texttt{SCPU} spends most of its time it the \texttt{train} function and
\texttt{ACPU} spends most of its time waiting during the training phase in
\Fref{fig:eg_tinn_net_train}.

\section{Conclusion} 
\label{sec:conclusion}
Although the experiment in \Fref{sec:probsys} may be a proof of concept example
the case of monitoring transactions on multiple communication channels to
ensure they are (or are not) cooperating is not uncommon where
traditional tools are often unable to provide useful visualizations.
Similarly, the experiment in \Fref{sec:tinn} represents a realistic use
case where an engineer is able to quickly, and with minimal manual effort, gain
useful knowledge about the workings of the software, without looking at
the source code.
These case studies and the example results shown in the figures demonstrate
that the methodology may be applied to a wide variety of situations in order
to aid the understanding and validation of complex systems.
Further work is being done with additional case studies
which combine hardware probes such as those in the probsys experiment with
software instrumentation such as that in the tinn experiment.

\bibliographystyle{ieeetr}
\bibliography{share/refs}{} 


\end{document}

%% file: share/tex/setupIEEE.tex
\input{share/tex/setup_misc.tex}
\input{share/tex/setup_math.tex}
\input{share/tex/setup_units.tex}
\input{share/tex/setup_listings.tex} 

\usepackage{cite} 

%% file: share/tex/setup_misc.tex
\def\BibTeX{{\rm B\kern-.05em{\sc i\kern-.025em b}\kern-.08em
    T\kern-.1667em\lower.7ex\hbox{E}\kern-.125emX}}

\usepackage[utf8]{inputenc}

\usepackage{pdflscape}

\usepackage{grffile} 

\ifdefined\NoGRAPHICX \else
\usepackage{graphicx}
\graphicspath{{./share/img/}}
\fi

\usepackage[plain]{fancyref}
\vrefwarning 

\usepackage{textcomp}

\usepackage{algorithmic}

\ifdefined\NoXCOLOR \else
\usepackage[usenames,svgnames]{xcolor}
\fi

\ifdefined\NoENUMITEM \else
\usepackage{enumitem}
\fi

\usepackage{placeins} 

\usepackage{multirow}
\usepackage{makecell} 

\ifdefined\NoTIKZ \else
\usepackage{tikz}
\usetikzlibrary{shapes, shapes.arrows}
\fi

\usepackage{glossaries}
\makenoidxglossaries
\input{share/tex/glossary.tex}

\usepackage{subcaption} 

\usepackage{breakurl}

%% file: share/tex/glossary.tex
\newglossaryentry{naiive}{
    name=na\"{\i}ve,
    description={is a French loanword (adjective, form of naïf)
        indicating having or showing a lack of experience, understanding or
        sophistication}
}

\newglossaryentry{ImageNet}{
    name={ImageNet},
    description={is an ongoing research effort to provide researchers around
        the world an easily accessible image database.}
}

\newglossaryentry{DutyFactor}{
    name={Duty Factor},
    description={Fraction of period where signal is high.}
}

\newacronym{ad}{AD}{Adaptive Design}
\newacronym{ai}{AI}{Artificial Intelligence}
\newacronym{amba}{AMBA}{Advanced Microcontroller Bus Architecture}
\newacronym{api}{API}{Application Prognamming Interface}
\newacronym{arm}{ARM}{Acorn RISC Machine Holdings Plc}
\newacronym{ascii}{ASCII}{American Standard Code for Information Interchange}
\newacronym{aws}{AWS}{Amazon Web Services}
\newacronym{axi}{AXI}{Advanced/ARM eXtensible Interface}
\newacronym{asic}{ASIC}{Application Specific Integrated Circuit}
\newacronym{bn}{BN}{Belief Network}
\newacronym{bnn}{BNN}{Binarized Neural Network}
\newacronym{bsc}{BSC}{Binary Symmetric Channel}
\newacronym{cam}{CAM}{Content Addressable Memory}
\newacronym{cdf}{CDF}{Cummulative Density Function}
\newacronym{cisc}{CISC}{Complex Instruction Set Computer}
\newacronym{cmos}{CMOS}{Complementary Metal Oxide Semiconductor}
\newacronym{cnn}{CNN}{Convolutional Neural Network}
\newacronym{cpd}{CPD}{Conditional Probability Distribution}
\newacronym{cpu}{CPU}{Central Processing Unit}
\newacronym{dag}{DAG}{Directed Acyclic Graph}
\newacronym{ddr}{DDR}{Double Data Rate}
\newacronym{dft}{DFT}{Discrete Fourier Transform}
\newacronym{dftst}{DFT}{Design For Testability}
\newacronym{Dkl}{$D_{KL}$}{Kullback Leibler Divergence}
\newacronym{dma}{DMA}{Direct Memory Access}
\newacronym{dsp}{DSP}{Digital Signal Processing}
\newacronym{dtft}{DTFT}{Discrete Time Fourier Transform}
\newacronym{ea}{EA}{Evolutionary Algorithm}
\newacronym{eeg}{EEG}{electroencephalograph}
\newacronym{evc}{EVC}{EVent Configuration}
\newacronym{evcx}{EVCX}{EVent Configuration eXpanded}
\newacronym{evs}{EVS}{EVent Samples}
\newacronym{fec}{FEC}{Forward Error Correction}
\newacronym{ff}{FF}{Flip-Flop}
\newacronym{fft}{FFT}{Fast Fourier Transform}
\newacronym{fifo}{FIFO}{First In First Out}
\newacronym{ops}{Op/s}{Operations Per Second}
\newacronym{fir}{FIR}{Finite Impulse Response}
\newacronym{foss}{FOSS}{Free Open Source Software}
\newacronym{fpga}{FPGA}{Field Programmable Gate Array}
\newacronym{ga}{GA}{Genetic Algorithm}
\newacronym{hdl}{HDL}{Hardware Description Language}
\newacronym{hls}{HLS}{High Level Synthesis}
\newacronym{huels}{HLS}{Hue-Lightness-Saturation}
\newacronym{hmm}{HMM}{Hidden Markov Model}
\newacronym{html}{HTML}{HyperText Markup Language}
\newacronym{ic}{IC}{Integrated Circuit}
\newacronym{iff}{iff}{if and only if}
\newacronym{iid}{IID}{Independent Identically Distributed}
\newacronym{iir}{IIR}{Infinite Impulse Response}
\newacronym{ip}{IP}{Intellectual Property}
\newacronym{iqr}{IQR}{Interquartile Range}
\newacronym{isa}{ISA}{Instruction Set Architecture}
\newacronym{jtag}{JTAG}{Joint Test Action Group}
\newacronym{kde}{KDE}{Kernel Density Estimation}
\newacronym{mac}{MAC}{Media Access Control}
\newacronym{macc}{MAcc}{Multiply-Accumulate}
\newacronym{md}{MarkDown}{MarkDown Markup Language}
\newacronym{mimo}{MIMO}{Multiple In, Multiple Out}
\newacronym{ml}{ML}{Machine Learning}
\newacronym{mm}{MM}{Minimax}
\newacronym{mmu}{MMU}{Memory Management Unit}
\newacronym{lcd}{LCD}{Liquid Crystal Display}
\newacronym{lfsr}{LFSR}{Linear Feedback Shift Register}
\newacronym{lha}{LHA}{Left Hand Associative}
\newacronym{lhs}{LHS}{Left Hand Side}
\newacronym{lrm}{LRM}{Language Reference Manual}
\newacronym{lut}{LUT}{Look Up Table Cell}
\newacronym{nn}{NN}{Neural Network}
\newacronym{ocp}{OCP}{Open Core Protocol}
\newacronym{od}{OD}{Original Design}
\newacronym{ooo}{OoO}{Out-of-Order}
\newacronym{pca}{PCA}{Principle Component Analysis}
\newacronym{pcb}{PCB}{Printed Circuit Board}
\newacronym{pdf}{PDF}{Probability Density Function}
\newacronym{ppa}{PPA}{Power, Performance, Area}
\newacronym{prng}{PRNG}{Pseudo-Random Number Generator}
\newacronym{ram}{RAM}{Random Access Memory}
\newacronym{rcg}{RCG}{Rocket Chip Generator}
\newacronym{rf}{RF}{Radio Frequency}
\newacronym{rgb}{RGB}{Red/Green/Blue}
\newacronym{rhs}{RHS}{Right Hand Side}
\newacronym{risc}{RISC}{Reduced Instruction Set Computer}
\newacronym{rnn}{RNN}{Recurrent Neural Network}
\newacronym{rtl}{RTL}{Register Transfer Language}
\newacronym{rv}{RISC-V}{RISC-Five}
\newacronym{sa}{SA}{Simulated Annealing}
\newacronym{sbsn}{SBSN}{Stochastic Bit-Stream Neuron}
\newacronym{sgb}{SGB}{Stochastic Gradient Boosting}
\newacronym{simd}{SIMD}{Single Instruction Multiple Data}
\newacronym{soc}{SoC}{System-on-Chip}
\newacronym{sql}{SQL}{Structured Query Language}
\newacronym{sv}{SV}{System Verilog}
\newacronym{svd}{SVD}{Singular Value Decomposition}
\newacronym{svg}{SVG}{Scalable Vector Graphics}
\newacronym{svhn}{SVHN}{Street View House Numbers}
\newacronym{svm}{SVM}{Support Vector Machine}
\newacronym{tl}{TileLink}{TileLink On-Chip Interconnect}
\newacronym{tlb}{TLB}{Translation Lookaside Buffer}
\newacronym{ts}{TS}{Time Series}
\newacronym{ua}{uarch}{Micro Architecture}
\newacronym{uob}{UoB}{University of Bristol}
\newacronym{usb}{USB}{Universal Serial Bus}
\newacronym{ust}{UltraSoC}{UltraSoC Technologies Ltd}
\newacronym{v95}{V95}{Verilog (1995)}
\newacronym{vcd}{VCD}{Value Change Dump}
\newacronym{vhdl}{VHDL}{Very High Speed Integrated Circuit Hardware Description Language}
\newacronym{vliw}{VLIW}{Very Long Instruction Word}
\newacronym{vlsi}{VLSI}{Very Large Scale Integration}
\newacronym{xml}{XML}{eXtensible Markup Language}
\newacronym{yaml}{YAML}{YAML Ain't Markup Language}

%% file: share/tex/setup_math.tex
\usepackage{amsmath,amssymb,amsfonts}
\usepackage{mathtools}
\usepackage{relsize} 

\newcommand{\indep}{\perp\!\!\!\perp} 
\DeclareMathOperator{\Ex}{\mathbb E} 
\DeclareMathOperator{\cov}{cov} 

\renewcommand{\leq}{\leqslant} 

\DeclareMathOperator{\sEx}{\mathbb E} 

\DeclareMathOperator{\sDep}{\dot{D}ep}
\DeclareMathOperator{\sCov}{\dot{C}ov}

%% file: share/tex/setup_units.tex
\usepackage[binary-units]{siunitx}

\DeclareSIUnit{\nothing}{\relax} 

\let\DeclareUSUnit\DeclareSIUnit

\DeclareUSUnit\ounce{oz} 
\DeclareUSUnit\inch{in} 
\DeclareUSUnit\foot{ft} 
\DeclareUSUnit\mil{mil} 

\DeclareSIUnit\dBi{dBi} 

%% file: share/tex/setup_listings.tex
\usepackage{listings}

\lstdefinelanguage{none}{
  identifierstyle=
}

\lstdefinelanguage{SystemVerilog}%
  {morekeywords={
      always_comb,always_ff,always_latch,%
      bit,logic,int,var,%
      always,and,assign,automatic,begin,buf,bufif0,bufif1,case,casex,%
      casez,cell,cmos,config,deassign,default,defparam,design,disable,%
      edge,else,end,endcase,endconfig,endfunction,endgenerate,%
      endmodule,endprimitive,endspecify,endtable,endtask,event,for,%
      force,forever,fork,function,generate,genvar,highz0,highz1,if,%
      ifnone,incdir,include,initial,inout,input,instance,integer,join,%
      large,liblist,library,localparam,macromodule,medium,module,nand,%
      negedge,nmos,nor,noshowcancelled,not,notif0,notif1,or,output,%
      parameter,pmos,posedge,primitive,pull0,pull1,pulldown,pullup,%
      pulsestyle_onevent,pulsestyle_ondetect,rcmos,real,realtime,reg,%
      release,repeat,rnmos,rpmos,rtran,rtranif0,rtranif1,scalared,%
      showcancelled,signed,small,specify,specparam,strong0,strong1,%
      supply0,supply1,table,task,time,tran,tranif0,tranif1,tri,tri0,%
      tri1,triand,trior,trireg,unsigned,use,vectored,wait,wand,weak0,%
      weak1,while,wire,wor,xnor,xor},%
   morekeywords=[2]{
      $clog2,$onehot,%
      $bitstoreal,$countdrivers,$display,$fclose,$fdisplay,$fmonitor,%
      $fopen,$fstrobe,$fwrite,$finish,$getpattern,$history,$incsave,%
      $input,$itor,$key,$list,$log,$monitor,$monitoroff,$monitoron,%
      $nokey},%
   morekeywords=[3]{
      `accelerate,`autoexpand_vectornets,`celldefine,`default_nettype,%
      `define,`else,`elsif,`endcelldefine,`endif,`endprotect,%
      `endprotected,`expand_vectornets,`ifdef,`ifndef,`include,%
      `no_accelerate,`noexpand_vectornets,`noremove_gatenames,%
      `nounconnected_drive,`protect,`protected,`remove_gatenames,%
      `remove_netnames,`resetall,`timescale,`unconnected_drive},%
   alsoletter=\`,%
   sensitive,%
   morecomment=[s]{/*}{*/},%
   morecomment=[l]//,
   morestring=[b]"%
  }[keywords,comments,strings]%

\newcommand*{\fancyreflstlabelprefix}{lst}

\fancyrefaddcaptions{english}{%
  \providecommand*{\freflstname}{listing}%
  \providecommand*{\Freflstname}{Listing}%
}

\frefformat{plain}{\fancyreflstlabelprefix}{\freflstname\fancyrefdefaultspacing#1}
\Frefformat{plain}{\fancyreflstlabelprefix}{\Freflstname\fancyrefdefaultspacing#1}

\frefformat{vario}{\fancyreflstlabelprefix}{%
  \freflstname\fancyrefdefaultspacing#1#3%
}

\Frefformat{vario}{\fancyreflstlabelprefix}{%
  \Freflstname\fancyrefdefaultspacing#1#3%
}